%Paper: hep-th/9505074
%From: David A. Lowe <lowe@tpau.physics.ucsb.edu>
%Date: Fri, 12 May 95 14:56:59 -0700
%Date (revised): Fri, 12 May 95 15:09:16 -0700

\input harvmac
\noblackbox

%
% Variable definitions
%
\def\frac#1#2{ {{#1} \over {#2}}}

\def\pmin{p^-}
\def\pp{p^+}

\def\xm{x^-}
\def\xp{x^+}

%
% Journal definitions
%
\def\CQG{{\it Class. Quan. Grav.\/}}

\def\NP{{\it Nucl. Phys.\ }}

\def\PL{{\it Phys. Lett.\ }}
\def\PR{{\it Phys. Rev.\ }}
\def\PRL{{\it Phys. Rev. Lett.\ }}
\def\CMP{{\it Comm. Math. Phys.\ }}

\def\IJMP{{\it Int. Jour. Mod. Phys.\ }}

\def\NC{{\it Nuovo Cimento \ }}

\font\tiau=cmcsc10
\baselineskip 12pt
\Title{\vbox{\baselineskip12pt \hbox{hep-th/9505074}
\hbox{UCSBTH-95-11} }}
{\vbox{\hbox{\centerline{\bf THE PLANCKIAN CONSPIRACY: STRING THEORY
AND}}
\hbox{\centerline{\bf THE BLACK HOLE INFORMATION PARADOX}}}}
\centerline{\tiau David A. Lowe\foot{lowe@tpau.physics.ucsb.edu}}
\vskip.1in
\centerline{\it Department of Physics}
\centerline{\it University of California}
\centerline{\it Santa Barbara, CA 93106-9530}

\vskip .5cm
\noindent
It has been argued that the consistency of quantum theory with
black hole physics requires nonlocality not
present in ordinary effective field theory. We examine the extent
to which such nonlocal effects show up in the perturbative S-matrix of string
theory.

\Date{May, 1995}

%
% References
%

\lref\swhI{S.~W.~Hawking, \PR {\bf D14} (1976) 2460.}
\lref\swhII{S.~W.~Hawking, \CMP {\bf 43} (1975) 199.}

% Susskind, Thorlacius, and Uglum, "The Stretched Horizon and Black
%Hole Complementarity"
\lref\stu{L.~Susskind, L.~Thorlacius, and J.~Uglum, \PR {\bf D48}
(1993) 3743.}

% 't Hooft
\lref\thooft{G.~'t~Hooft, \NP {\bf B335} (1990) 138;
{\it Phys. Scr.} {\bf T36} (1991) 247, and references therein.}

% Schoutens, Verlinde, and Verlinde
\lref\svv{E.~Verlinde and H.~Verlinde, \NP {\bf B406} (1993) 43;
K.~Schoutens, E.~Verlinde, and H.~Verlinde \PR {\bf D48} (1993)
2690; Y.~Kiem, H.~Verlinde and E.~Verlinde, hep-th/9502074.}

\lref\lenny{L.~Susskind, \PRL {\bf 71} (1993) 2367;
\PR {\bf D49} (1994) 6606.}

\lref\gedan{L.~Susskind and L.~Thorlacius, \PR {\bf D49} (1994) 966.}

% Martinec, "The Light Cone in String Theory", "Strings and
%Causality"
\lref\martinec{E.~Martinec \CQG {\bf 10} (1993) L187;
{\it Strings and Causality,} preprint EFI-93-65, hep-th/9311129.}

% Lowe, "Causal Properties of Free String Field Theory"
\lref\lowe{D.~A.~Lowe, \PL {\bf B326} (1994) 223.}

\lref\lsu{D.~A.~Lowe, L.~Susskind, and J.~Uglum, \PL {\bf B327}
(1994) 226.}

%Klebanov et al
\lref\kleb{M.~Karliner, I.~Klebanov and L.~Susskind, \IJMP {\bf A3}
(1988) 1981; A.~Mezhlumian, A.~Peet and L.~Thorlacius, \PR {\bf D50}
(1994) 2725.}

% Susskind, "The World as a Hologram"
\lref\holo{L.~Susskind, {\it The World as a Hologram, \/} Stanford
University preprint SU-ITP-94-33, September 1994, hep-th/9409089.}

%The big black hole paper
\lref\binfo{D.~A.~Lowe, J.~Polchinski, L.~Susskind, L.~Thorlacius and
J.~R.~Uglum, to appear.}

%\draftmode
\newsec{Introduction}

The problem of information loss during quantum black hole evaporation
\refs{\swhI,\swhII}
provides us with an important paradox which is likely to point
the way to a better understanding of quantum gravity.
If one takes the point of view that quantum mechanics is not
drastically modified, an evaporating black hole must either: (i)
encode the information carried by matter falling through the event horizon
in the outgoing Hawking radiation and at the same time
leave a freely falling observer crossing the event horizon
essentially undisturbed \refs{\stu,\thooft,\svv},
or (ii) store the information in a long-lived or stable remnant
\ref\aharon{Y.~Aharonov, A.~Casher and S.~Nussinov, \PL {\bf 191B} (1987) 51.}.
The second possibility suffers
from the serious drawback that such remnants are likely to be
infinitely produced in ordinary low energy scattering experiments
\ref\giddings{S.~B.~Giddings, {\it Why aren't black holes infinitely
produced?}, hep-th/9412159, and references therein.}.
The studies in two dimensions of semiclassical dilaton-gravity models
\ref\twod{C. G. Callan, Jr., S. B. Giddings, J. A. Harvey and A. Strominger,
\PR {\bf D45}, (1992) R1005; S. W. Hawking, \PRL {\bf 69} (1992) 406;
B. Birnir, S. B. Giddings, J. A. Harvey and A. Strominger, \PR {\bf D46}
(1992) 638; T. Banks, A. Dabholkar, M.R. Douglas and M. O' Loughlin,
\PR  {\bf D45} (1992) 3607;
J.G. Russo, L. Susskind and L. Thorlacius, \PL {\bf B292} (1992) 13,
\PR {\bf D46} (1992) 3444, \PR {\bf D47} (1993) 533;
A.~Bilal and C.~G.~Callan, \NP {\bf B394} (1993) 73;
S.~P.~de~Alwis, \PL {\bf 289} (1992) 278;
D. A. Lowe, \PR {\bf D47}, (1993) 2446.}
provide strong evidence that the first possibility is not realized
in the context of local field theoretic degrees of freedom coupled
to gravity.

For a large mass black hole, all local invariants outside and in the vicinity
of the horizon are small for a large fraction
of the lifetime of the black hole. In this case one might expect to
be able to set up a local low energy effective field theory on this
slowly varying spacetime, on a family of suitably defined slices.
Further details of this argument, which we
refer to as the nice slice argument, may be found in \binfo.
The usual rules of local quantum field theory tell us that
local operators inside commute with local
operators outside the horizon.
Since the fields inside the horizon are correlated with the
infalling state, an observer outside will conclude that information has
been lost.

This argument suggests that if information is to be encoded
in Hawking radiation, some new kind of nonlocality not present
in local effective field theory is required.
At the same time,
these new nonlocal effects should not spoil ordinary low energy physics
when black holes are not present. In this paper we further explore the
idea that string theory realizes this possibility. This point of view
has been considered previously in \refs{\lenny,\gedan}.

Because strings are extended objects, they are sensitive to
nonlocal geometric invariants. This provides the needed loophole
in the nice slice argument.
In the black hole problem there is a large nonlocal invariant -- the
relative boost between the infalling and outgoing frame in which
Hawking radiation is measured.
We will show this leads to nonlocal effects over macroscopic
spacetime separations visible in string S-matrix elements.
We argue that these effects
undermine the usual nice slice argument and that a nonlocal effective
field theory is necessary to correctly describe low energy physics
on a family of nice slices.

Another approach to this problem would be to compute the commutator
of string fields in some suitable string field theory.
In \refs{\martinec,\lowe}, it was argued that the  commutator of
free string fields in light-cone gauge was essentially local.
This is no longer the case in the interacting theory \lsu, where
the so-called transverse spreading was studied.
In \binfo, this result is generalized to string fields with nontrivial
separation in the longitudinal plane.
There the nonlocal effect found is much larger than that
present in the S-matrix. However, then one is faced with the
difficult question of the interpretation of these
off-shell quantities. It is possible these large nonlocal effects
are simply a gauge artifact. On the other hand, these off-shell
effects in flat space could show up in on-shell physical amplitudes
computed  around a finite mass black hole type background.
Regardless of which interpretation turns out to be correct, the
nonlocal behavior present in the S-matrix should be thought of as a
lower bound on the degree of nonlocality present in string theory.\foot{This
of course assumes that we are treating the string degrees of freedom as
fundamental. It remains a logical possibility that string theory
may be formulated in terms of an underlying local theory.}

The paper is organized as follows. In section 2, tree-level
S-matrix elements corresponding to the interaction of Hawking type
particles with a particle freely falling across an event horizon
are considered. These amplitudes fall off over a characteristic
length scale that becomes macroscopically large at sufficiently late times.
In section 3, we consider some of the physical consequences of these
amplitudes and argue that they support the notion that degrees
of freedom on a stretched horizon (a timelike surface lying outside the
true event horizon) retain information about the infalling state.
Section 4 discusses higher genus corrections to tree-level results,
and we conclude with some brief remarks in section 5.

\newsec{String Spreading in the Regge Limit}

In the following we will be interested in computing string S-matrix
elements between a set of operators at low energy relative
to an observer freely falling into a large mass $M_{bh}$ black hole,
and another set of operators at low energy relative to an observer
who stays outside the black hole and measures the Hawking radiation at late
times.  We will refer to such operators as ``nice'' operators.
The stringy effects of interest here survive as
the mass of the black hole goes to infinity, at which point the
region of spacetime near the horizon is simply flat space.
It suffices, therefore, to compute the string amplitudes in flat
space, folding in appropriate wavepackets to construct the
operators of interest. We denote by $a$ the relative boost factor
between the two frames. For a Schwarzschild background this
increases exponentially with Schwarzschild time $t_S$, $a= e^{t_S/4M_{bh}}$,
becoming
very large at late times. We use the null coordinates $X^+$, $X^-$
to parametrize the longitudinal plane of the spacetime. The horizon
corresponds to the line $X^+ = 0$, with $X^+ < 0$ outside the horizon.
The $D-2$ transverse coordinates are denoted $\vec X$ and the
conventions for the metric are $ds^2 = 2 dX^+ dX^- - (d \vec X)^2$.

With respect to these Minkowski coordinates, the center of mass energy
squared $s$ of these amplitudes is very large at late times (of order $a$),
while the momentum transfer squared $t$ is small. In this limit the Regge
behavior of the string amplitudes is apparent.  Regge-Gribov
techniques \ref\rgrib{V.N. Gribov, {\it Sov. Phys. JETP} {\bf 26}
(1968) 414.} may be used to compute the string amplitudes.
This procedure has been considered previously in
\ref\amati{D. Amati, M. Ciafaloni and G. Veneziano,
\PL {\bf 197B} (1987) 81; \IJMP {\bf A3} (1988)
1615; \PL {\bf 216B} (1989) 41.}. The amplitudes are factorized onto fictitious
Reggeon particles with angular momentum $J = \alpha(t) = 2+\alpha' t/2$.
Vertices coupling these Reggeons to ordinary particles such as gravitons,
may be extracted by factorizing tree-level amplitudes. These vertices
may then be sewn together to yield amplitudes at arbitrary numbers of
loops.

We will adopt a simpler method, approximating the string
path integral by a saddle-point calculation. This gives the correct
asymptotic dependence on $s$ for the amplitudes of interest here,
but in general will not yield the
full $t$ dependence. Fortunately, the full $t$ dependence will not
be important in the following.
The saddle-point approximation to the string path integral has been
considered before by Gross and Mende
\ref\gross{D.J. Gross and P.F. Mende, \PL {\bf 197B} (1987) 129;
\NP {\bf B303} (1988) 407.}, in the context of large $s$,
fixed angle (i.e. fixed $t/s$) scattering. Taking the fixed $t$ limit
of their expressions yields the amplitudes of interest here.
It may be shown that when the determinant factors from the integration
over fluctuations in the moduli are included, with the appropriate measures,
the large $s$ asymptotics coincide with those obtained via Regge-Gribov
methods, despite the fact that for fixed $t< 1/\alpha'$ the saddle-point
approximation is no longer strictly valid.

The string path integral is
\eqn\spath{
A = \int {\cal D} X \exp( -{1\over {4\pi \alpha'}} \int \del X \bar \del X
+ i \sum_j p_j \cdot X(z_j) ) ~.
}
The saddle point is dominated by the worldsheet
\eqn\worlds{
X^{\mu}_{cl} = i \alpha' \sum_j p^{\mu}_j \log|z-z_j| ~,
}
which leads to the amplitude
\eqn\samp{
A \sim \exp(- \alpha' E)~,
}
where $E$ may be thought of as the electrostatic energy of charges
$p_j^{\mu}$ on the worldsheet at positions $z_j$
\eqn\electroen{
E=  \sum_{i<j} p_i \cdot p_j \log|z_i - z_j|~.
}
This energy is then to be minimized by varying the independent
moduli.

As an example, consider the four-graviton scattering amplitude at
tree-level. The mass-shell condition leads to the relation
$s+t+u =0$, where $s,t,u$ are the Mandelstam variables:
 $s=(p_1+p_2)^2$, $t= (p_1+p_3)^2$ and $u=(p_1+p_4)^2$.
The energy $E$ is invariant under $SL(2,C)$ transformations
at tree-level, so depends on only one complex modulus
\eqn\imodu{
x= { { (z_1-z_3) (z_2-z_4)}\over { (z_1 -z_2) (z_3 - z_4) }} ~.
}
Equation \electroen\ then is rewritten
\eqn\fgen{
E= {1\over 2} \bigl( t \log |x| + u \log|1-x| \bigr) ~.
}
Minimizing with respect to $x$ then gives the solution
\eqn\mmod{
x_{cl} = - t/s~,
}
so that in the limit $t \ll s$, $E \sim - {1\over 2} t \log s$.
Note that because $x_{cl} \ll 1$ in this limit, the amplitude
factorizes. This kind of factorization between
the infalling operators and the outgoing Hawking operators will
be a property of the generic $N$ particle amplitudes.

Including the determinant from fluctuations in $x$, with the appropriate
measure, the final amplitude is
\eqn\fgravamp{
A(s,t) \sim s^{2 + \alpha' t/2}~.
}
This amplitude could not be obtained from a local field theory
with a finite number of fields, since \fgravamp\ violates
polynomial boundedness. There one would obtain at tree-level
an amplitude polynomial in the momenta, multiplied by factors containing
poles coming from internal propagators.
One may then view this amplitude \fgravamp\ as a field theory amplitude
multiplied by a form factor corresponding to the effective size
of the strings,
\eqn\fgravf{
A(s,t) \sim s^2 F^2_s(t)~,
}
where the form factor is
\eqn\formfac{
F_s(t) = \exp( {{\alpha' t} \over 4} \log s)~.
}
This tells us the effective size of the strings is given by
the Lorentz invariant expression
\eqn\effecsiz{
(\delta X^\mu)^2 \sim \alpha' \log s~.
}

An important point to note is that if we consider the four-point
amplitude of a nice low energy pair of Hawking particles
$(\pp \sim 1/a\ell,~\pmin \sim a/\ell )$ in one frame
and a nice low energy pair of infalling operators
$(\pp \sim 1/\ell,~\pmin \sim 1/\ell)$ in the other frame,
kinematics restricts the momentum transfer to be purely transverse,
in the large $s$ limit. Here $\ell$ is defined to be some typical
experimental scale.
This means the amplitude is not sensitive
to the longitudinal spreading indicated in \effecsiz, i.e.
$\delta X^+ \delta X^- \sim \alpha' \log s$.
This turns
out to be a general feature of the Regge amplitudes of fixed numbers
of nice particles in the large $s$ limit.

One way
to ensure that the longitudinal momentum transfer $q^+ q^-$ is comparable
to $\vec q^2$, is to consider amplitudes with large numbers ($O(a)$)
of nice operators.
However, the extra large number in the problem generally
leads to the break down of the above approximations. Alternatively one
can consider amplitudes involving states with $M^2 \sim a$. These
states may be thought of as long strings stretched between the Hawking
probe and the infalling operators. These
amplitudes can be computed within the saddle-point approximation,
and do exhibit the longitudinal spreading indicated by \effecsiz, as
will be shown in the following.

To simplify matters further, we will replace the large  $M^2$ state
by two graviton fields with momenta that are large with respect to
both the infalling and Hawking frames,
and consider the five-graviton amplitude corresponding to the
production of these gravitons via a pair of Hawking operators interacting
with an infalling operator.
These two graviton fields are not ``nice'' operators, according
to our previous definition, but nevertheless they may be produced
via the interaction of some set of nice fields.
This corresponds to an experiment
in which the Hawking observer conducts a scattering experiment at low
energy in her frame which involves a high-energy interaction
with the infalling operator,
which produces this high-mass intermediate state. The high-mass state
then decays into a pair of high-energy gravitons. Of course this
is not the most likely final state, which is expected to involve
a large number of particles. We should not be surprised then if this
amplitude is highly suppressed -- for now we are only interested in whether
the longitudinal string spreading suggested by \effecsiz\ shows up.

The amplitude is obtained by minimizing \electroen\ with respect to
the two independent complex moduli. Let us denote the momenta of the
Hawking particles by the subscripts $h$ and $h'$, those of the
infalling particle by $i$, and the momenta of the outgoing pair of
gravitons $j$ and $j'$. The products of momenta satisfy:
$p_h \cdot p_{h'} < O(1)$, with all other products between different
momenta of order $a  \gg 1$. The equations to be solved are
\eqn\nmin{
\sum_{k \neq l} { {p_l \cdot p_k } \over {z_l - z_k}} = 0~,
}
with no sum on $l$.
Up to $SL(2,C)$ transformations, given the above conditions on the
products of momenta, the solution satisfies
$|z_h - z_{h'}| \sim O(1)$, with all other $|\delta z|$ of order $a$.
This indicates the Hawking operators factorize with the exchange of a
Reggeon coupling them to the other operators.
Computing the amplitude leads to
\eqn\namp{
A(s,t,M^2,\phi) \sim s^{ \alpha' t/2} e^{-\alpha' M^2 f(\phi)/2} ~,
}
neglecting additional analytic factors from the measure and fluctuations in the
moduli. Here $s$ is of order $a$, $t=(p_h + p_{h'})^2$,
$M^2 = (p_j + p_{j'})^2$, and $f(\phi)$ is a function of the scattering angle
$\phi$ defined by $\sin^2 (\phi/2) = -2 p_i.p_j/M^2$ which has the form
\eqn\fis{
f(\phi) = - \bigl( \sin^2(\phi/2) \log \sin^2(\phi/2) +
\cos^2(\phi/2) \log \cos^2(\phi/2) \bigr)~.
}
The function $f(\phi)$ will generically be of order one.

Again we see the appearance in \namp\ of the $s^{\alpha' t/4}$
form factor which
is present whenever Reggeon exchange dominates the amplitude.
The longitudinal
components of the momentum transfer satisfy $q^+ \sim 1/a\ell$,
$q^- \sim a/\ell$
so that $q^+ q^- \sim O(\vec q^2)$.
Therefore, the momentum transfer is no longer purely transverse as in
the usual Regge amplitudes, and the
amplitude displays the longitudinal spreading $\delta X^+ \delta X^-
\sim \alpha' \log s$ implied by equation \effecsiz. This will
be a general property of diffractive scattering amplitudes
of sets of nice operators which lead to a long string (i.e.
a state with $M^2 \sim a$) in the final state.

Because this amplitude is suppressed by the factor
$\exp( -\alpha' M^2 f(\phi)/2)$, one might worry that typical
string configurations are not spread in this manner.
If one considers inclusive amplitudes in which the different final states
are summed over, the exponentially large number of final states with
center of mass energy $M$ will at least partially compensate for this
suppression factor. Nevertheless, the suppression factor is independent
of $t$, so we may fix $M^2$, $\phi$ and vary $t$ independently. The length
scale over which the amplitude falls off as we vary $t$ gives
a measure of the typical longitudinal component of the spreading.
Of course, on the basis of \effecsiz, one could argue such spreading
simply follows from Lorentz invariance. The purpose of the above
calculation is to demonstrate such spreading is in fact manifest in
S-matrix elements.

Let us
now consider further the spacetime properties of the form factor \formfac.
The Fourier transform of the form factor yields
the density of the extended object
\eqn\dense{
\rho( \delta x) = \int dq e^{i q \cdot \delta x} F_s(q^2)~.
}
Substituting \formfac\ into \dense, implies that the typical
size of the string is \effecsiz. In light-cone gauge this yields
the usual $\log s$ transverse spreading \refs{\lenny,\kleb,\lsu}.
Using \dense\ to
compute the longitudinal spreading leads to
\eqn\longsp{
\vev{ \delta X^-} \sim \alpha' \vev{q^-} \log s, \qquad
\vev{ \delta X^+} \sim \alpha' \vev{q^+} \log s~,
}
where $\vev{q}$ is the typical momentum transfer of the string
under consideration.
This is similar to the expression derived in \lenny\
but contains an additional
subleading $\log s$ factor. This implies that a string stops Lorentz
contracting as its apparent length reaches the string scale, and actually
begins growing logarithmically with the boost. These expressions should
be contrasted with the longitudinal spreading one finds in the
off-shell commutator in light-cone gauge. The power law fall-off with
$X^-$ in that case implies no length scale is associated with the
longitudinal spreading. See \binfo\ for further details.

If one folds in wavepackets with a typical spread $\ell$ (in space and
time, for example), and we work in the center of mass frame of the string,
a stringy uncertainty relationship is obtained
\eqn\suncert{
\delta X^{\mu} \sim {{\alpha'} \over \ell} \log s + \ell~.
}
This implies there is a minimum observable length scale obtainable
in performing fixed momentum transfer, high energy scattering
experiments
\eqn\xmin{
\delta X^{\mu}_{min} \sim \sqrt{\alpha' \log s}~.
}
Similar uncertainty relationships have been considered before in
\refs{\amati,\gross}.

\newsec{Physical Consequences}

Consider an accelerating observer who hovers outside the
horizon of a large mass
$M_{bh}$ black hole, such that she undergoes a proper acceleration
$1/\rho_0$, and performs experiments on length scale $\ell$.
The path the observer follows is then
\eqn\opath{
\xm = \rho_0 e^{t_S/4g^2 M_{bh}} , \qquad \xp = -\rho_0 e^{-t_S/4 g^2 M_{bh}}~,
}
where $t_S$ is the time measured by the observer and $g$ is the
string coupling constant.
We take the infalling operators to be localized near $\xm = 0$.
As argued above, including the effects
of the finite size of wavepackets, the infalling and Hawking operators
will interact over a range
\eqn\larcom{
\xm < \alpha' \vev{q^-} \log a + 1/\vev{q^+}=
 {{\alpha' t_S}\over {4 g^2 M_{bh} \ell}} e^{t_S/4 g^2 M_{bh}} + \ell e^{t_S/4
g^2 M_{bh}}
{}~,}
or equivalently
\eqn\comcond{
\rho_0 < {{\alpha' t_S}\over {4 g^2 M_{bh} \ell}}+ \ell~.
}
The first term in \comcond\ is the inherently stringy spreading, while the
second term is just the usual field theoretic spreading, arising from
the finite size of the detector.
In principle, $\ell$ can be made as small as one
wishes so that this field theoretic overlap may be neglected. However,
as $\ell$ is made very small, the effective size of the string
increases. One concludes the nonlocal terms are always significant when
\eqn\cnonv{
\rho_0 < \sqrt{ {{\alpha' t_S}\over {4 g^2 M_{bh}} }}~.
}

If we choose to view string theory in terms of a set of degrees
of freedom local with respect to some fixed target space background,
the above result suggests that degrees of freedom
on a stringy stretched horizon $\xp \xm = \alpha'\log(a)$
carry information about the infalling state.
The size of this horizon differs by a factor $\log(a)$ with
the analogous result of \lenny.

Note the above calculations were based on perturbative string
theory. As we will see later,
in the fixed momentum transfer, large $s$ limit strong coupling effects
become important when
$s > 1/( \alpha' g^2)$, where $g$ is the string coupling constant.
We conclude the above result \cnonv\ is only applicable for time
\eqn\vtime{
t < O( 4 M_{bh} g^2 \log 1/g^2 ).
}

A detector must either
undergo string scale accelerations or cross over the path of
the infalling body, to attempt to detect these
degrees of freedom with this kind of experiment.\foot{This assumes $\log 1/g^2$
is roughly of order 1. If $\log 1/g^2 \gg 1$ then the string spreading
effects come into play at substring-scale accelerations.}
This does not directly lead to any breakdown
of low energy effective field theory, unless we are prepared
to extrapolate the tree-level results beyond their range of validity \vtime.
However, in the case of a
finite mass black hole, a similar stretched horizon will be present,
which couples to low energy observers via Hawking radiation. It seems
plausible that these information carrying degrees of freedom
on the stretched horizon will couple to the Hawking modes, allowing
information to leak off the stretched horizon. In terms of a low energy
effective field theory on a set of nice slices, these effects will
show up as nonlocal interactions between the Hawking region and the
infalling region.

On the other hand,
since string theory appears to allow us to ask questions involving
infalling and outgoing operators at the same time, if we assume the picture
suggested by above is valid at very late times, there
would appear to be a problem with large scale violations of the
equivalence principle. Any observation on the Hawking radiation
which yields information about the state fallen into the black hole,
would seem to lead to a violent perturbation of the infalling
observer as she passes through the horizon.

Of course, the fact that the amplitude \namp\ is highly suppressed
means such processes occur extremely rarely. Consequently, these
processes can transfer only a small amount of information to the
outgoing Hawking radiation. However, we regard the existence
of these rare information transferring perturbative processes as an
important hint that string theory may realize possibility (i) of the
Introduction.

Clearly something more subtle must happen if information retrieval
before the endpoint of black hole evaporation
is to be implemented in string theory.
It is possible the new physics enters in at the nonperturbative level. The
perturbative picture described above, in which stringy
degrees of freedom carry information on the stretched horizon about
the infallen state, will only provide a correct description of the
physics for sufficiently early times \vtime. After that time,
much larger effects must set in if all the
information of the infalling state is to eventually reach an
outside observer in the form of Hawking radiation. To be
consistent with known physics, the infalling observer cannot
perceive these effects until she approaches the singularity.
This suggests that a better understanding of the stringy
physics near the singularity is required to determine whether all
information escapes encoded in the Hawking radiation.

\newsec{Extension to Higher Genus}

At arbitrary genus $G$, a saddle-point analysis similar to the above
gives an estimate of the large $s$ asymptotics of the fixed $t$ amplitude.
Let us restrict attention to the genus $G$ four-graviton amplitude.
This is dominated by a term which comes from sewing together
tree-level amplitudes
\eqn\gamp{
A_G(s,t)  \sim \prod_{i=1}^{G+1} A_{{\rm tree}} (s,t_i)~,
}
which is maximized subject to the condition that
$\sum \sqrt{-t_i} \geq \sqrt{-t} $. It follows from the
form of the tree-level amplitudes that the maximum occurs when
the momentum transfer is equally shared between the different
internal legs, so that $\sqrt{-t_i} \sim \sqrt{-t}/(G+1)$.
When the determinants from the fluctuations of the
moduli with the correct measures are included \gross, one finds
\eqn\gampl{
A_G(s,t) \sim (g^2 s)^{G+1} i^G (G+1)^{9G} s^{ \alpha' {t\over {2(G+1)}} +1}~.
}
Neglecting the effect of the $(G+1)^{9G}$ factor for the moment, we see
that $g^2 s$ is an effective expansion parameter in the large $s$, fixed
$t$ regime, and we should expect strong coupling effects to become
important when $\alpha' s > 1/g^2$.

Taking into account the $(G+1)^{9G}$ behavior, the sum of the leading terms
\eqn\persum{
A_{\rm sum}(s,t) \sim \sum_{G=0}^{\infty} A_G(s,t)
{}~,}
will diverge for fixed $s$, $t$ and $g$. However,
it may be Borel resummed along the lines of
\ref\ooguri{P. Mende and H. Ooguri, \NP {\bf B339} (1990) 641.}, to yield
a convergent expression with the same asymptotic expansion in $g$ as \persum
\eqn\resum{
A_{\rm resum}(s,t) = \int_0^{\infty} dz e^{-z} \sum_{G=0}^{\infty}
{z^{9N}\over (9N)!} A_G(s,t) ~.
}

We may evaluate the Borel resummation \resum\ using
a saddle-point approximation \ooguri\ which yields
\eqn\sumans{
|A_{\rm resum}(s,t) |\sim e^{- \sqrt{-2 \alpha' t \log s \log(1/g^2) } }
{}~.}
The value of $G$ at the saddle point is
\eqn\gbar{
G_0+1 = \sqrt{ {{-\alpha' t \log s} \over {2 \log(1/g^2)}} }
{}~,}
which must be large ($G_0 \gg 1$) for the saddle-point approximation
to be valid. Equation \sumans\ then is valid when
\eqn\inrang{
{-2\over \alpha' t} \log (1/g^2) \ll \log s~.
}
If ${-2\over \alpha' t} \log (1/g^2) \gg \log s$, the tree-level
term will dominate the series, which leads to the
$(\delta X)^2 \sim \alpha' \log s$ spreading described in the previous
sections.

Since we have summed only the leading terms, one might
worry that subleading terms might change the asymptotic
behavior given by \sumans. Subleading corrections may be estimated
and resummed using the above procedure as described
in \ooguri. The dominant corrections
appear to come from integrals over the moduli and take the
form $G^3/t ~A_G(s,t)$. In the resummed amplitude, these will
be negligible if $G_0^3/t \ll 1$ which yields the additional
constraint $ |\alpha' t/2|^{1/3} \log s \ll \log(1/g^2)$. Therefore the
above approximation to the resummed amplitude \sumans\ is at best only valid
in the range
\eqn\rang{
({1\over {g^2}} )^{2/\alpha' |t|} \ll \alpha' s \ll
({1\over {g^2}} )^{(2/\alpha' |t|)^{1/3}}~.
}
This region in parameter space
will only exist when $|t| > \alpha'/2$, so this is telling us about
the dense central region of string that appears as the outer extremities
undergo the logarithmic spreading described by the tree-level term \kleb. Of
course, it has not been proven that the subleading
terms are negligible when all subleading corrections are
taken into account. It remains a possibility that these terms conspire
to eliminate the range of validity of \sumans.

For now, let us assume the resummed amplitude is valid for some
range with $\log s > {2\over \alpha' |t|} \log (1/g^2)$.
Equation \sumans\ may then be interpreted as a product of form factors,
as in \fgravf, now with higher genus corrections included.
The resummed form factor falls off more slowly for large $|t|$, than
the tree-level result. This slower
falloff turns out to be consistent with the bound of Cerulus and Martin
\ref\cerulus{F. Cerulus and A. Martin, \PL {\bf 8} (1963) 80;
A. Martin, \NC {\bf 37} (1965) 671.} on the fastest possible
falloff in a local field theory, indicating that the resummed amplitude
behaves in a more local way than just the tree-level term alone.
Consequently, strings are more dense in the central region
than would be expected from the tree-level result. Arguments
presented in \holo\ suggest this behavior does not
persist for $\alpha' s > 1/g^2$.
There it was argued a much faster
falloff must appear if string theory is to be consistent
with the Beckenstein bound on the maximum amount of information
within a fixed volume.

\newsec{Conclusions}

In this paper, we have studied certain perturbative string S-matrix elements
relevant to the black hole information problem.
These matrix elements correspond to rare
events which transfer information from an infalling body to
the Hawking radiation in a nonlocal manner,
disturbing the infalling body in the process.
Such processes do not appear in a local effective field theory
which might be expected to describe low energy physics
on a family of nice slices of a large mass black hole.
We therefore regard the existence of such processes in string theory
as evidence that the effective field theory must contain nonlocal
interactions. This nonlocality undermines the usual argument for
information loss.

The perturbative results we have found are only valid for
sufficiently early times, so cannot be the whole story.
It is possible nonperturbative effects play an important role at later
times. These effects must be much larger than the perturbative
effects described above, if all infalling information is to escape in the
form of Hawking radiation.

We have argued that string theory supports the idea that
degrees of freedom on a stretched horizon retain information
about what fell into the black hole. These degrees of freedom
may then be interpreted as a kind of stringy hair. This
is allowed, because when fields with spin greater than 2 are present,
the usual no hair theorem of classical general relativity
breaks down. Such hair is usually associated with some conserved
quantity. It would be interesting to identify the
symmetries of string theory which lead to these conserved quantities.

\vskip .5in
{\bf Acknowledgements}

It is a pleasure to thank S.~Giddings, G.~Horowitz, J.~Polchinski,
M.~Srednicki, A.~Strominger,
L.~Susskind, L.~Thorlacius and J.~Uglum for helpful discussions and comments.
I also thank J.~Polchinski, L.~Susskind, L.~Thorlacius and J.~Uglum
for collaboration on the related work \binfo.
This work was supported in part by NSF grant PHY-91-16964.

\listrefs
\end